\documentclass[twocolumn,pre,showpacs,amsmath,amssymb,superscriptaddress]{revtex4}

\usepackage{graphicx}
\usepackage{bm}
\usepackage{color}

\newcommand{\bS}{{\bf s}}

\begin{document}
\title{The Ising Model for Neural Data: Model Quality
and Approximate Methods for Extracting Functional Connectivity}
\author{Yasser Roudi}
\affiliation{NORDITA, Roslagstullsbacken 23, 10691 Stockholm, Sweden}
\author{Joanna Tyrcha}
\affiliation{Dept of Mathematical Statistics, Stockholm University, 10691 Stockholm, Sweden}
\author{John Hertz}
\affiliation{NORDITA, Roslagstullsbacken 23, 10691 Stockholm, Sweden}\affiliation{The Niels Bohr Institute, Copenhagen University, 2100 Copenhagen \O, Denmark}

\begin{abstract}
We study pairwise Ising models for describing the statistics of
multi-neuron spike trains, using data from a simulated cortical
network. We explore efficient ways of finding the optimal couplings
in these models and examine their statistical properties. To do
this, we extract the optimal couplings for subsets of size up to
$200$ neurons, essentially exactly, using Boltzmann learning. We
then study the quality of several approximate methods for finding
the couplings by comparing their results with those found from
Boltzmann learning. Two of these methods -- inversion of the TAP
equations and an approximation proposed by Sessak and Monasson --
are remarkably accurate. Using these approximations for larger
subsets of neurons, we find that extracting couplings using data
from a subset smaller than the full network tends systematically to
overestimate their magnitude.  This effect is described
qualitatively by infinite-range spin glass theory for the normal 
phase. We also show that a globally-correlated
input to the neurons in the network lead to a small increase in the
average coupling. However, the pair-to-pair variation of the
couplings is much larger than this and reflects intrinsic properties
of the network. Finally, we study the quality of these models by
comparing their entropies with that of the data.  We find that they
perform well for small subsets of the neurons in the network, but
the fit quality starts to deteriorate as the subset size grows,
signalling the need to include higher order correlations to describe
the statistics of large networks.
\end{abstract}


\maketitle

\section{Introduction}

Computation in the brain is performed by large populations of
neurons. Because these neurons are highly-connected, and because the
external inputs that they receive is usually correlated, the
neuronal spike trains are also correlated. These correlations depend
on neuronal properties, synaptic connectivity and the external drive
in a highly nontrivial way.

The large number of neurons involved in any computation, and the
fact that they are correlated, make deciphering the mechanisms of
neural computations a difficult challenge. A major technical
breakthrough in this challenge has been the advent of techniques for
recording simultaneously from large numbers of neurons. Yet, making
a link between these recordings and an understanding of the
computations is nontrivial and requires new mathematical
approaches. This is because, in most cases, using the observed data
to answer questions about computation requires building statistical
models, i.e. writing down the probability distribution over spike
patterns \cite{Rieke97}. However, the high dimensionality of the
space of possible spike patterns makes it very hard to collect
enough data to build an exact statistical description of them.

One approach to circumvent the problem of high dimensionality of the
space of spike patterns is to use parametric models. In this
approach, one uses the data to fit a parametric probability with a
much smaller number of parameters than the dimensionality of the
space of spike patterns. To use any such parametric model reliably,
one needs to answer two questions: How can we fit the parameters of
the model efficiently, and how close is the model to the true
probability distribution?

In this paper, we try to provide answers to these questions for the
case of the maximum entropy binary pairwise model, the Ising model,
using data from a simulated network of spiking neurons. The Ising
model has received a lot of attention as a parametric model for
neural data following the study by Schneidman {\em et al}
\cite{Schneidman05}. These authors modeled the true distribution
of the spike patterns by the Gibbs distribution of an Ising model:
\begin{equation}
p_{\rm Ising}(\bS)\propto \exp\left\{\sum_{i}h_i s_i +\sum_{i<j}
J_{ij} s_i s_j\right\}, \label{p-ising}
\end{equation}
where $\bS=(s_1,s_2,\dots,s_N)$, and each spin $s_i=\pm 1$
represents the firing or not firing of neuron $i$. The external
fields, $h_i$, and the coupling parameters, $J_{ij}$, were fit so
that the resulting distribution had the same means and pairwise
correlations as the data, that is,
\begin{subequations}
\begin{eqnarray}
&\langle s_i \rangle_{\rm Ising}=\langle s_i \rangle_{\rm data}
\label{means-p-data}\\
&\langle s_i s_j \rangle_{\rm Ising}=\langle s_i s_j \rangle_{\rm
data}\label{corrs-p-data},
\end{eqnarray}\label{means-corrs-p-data}
\end{subequations}where $\langle \rangle_{\rm Ising}$ represents averaging with
respect to the Ising model distribution (\ref{p-ising}), and $
\langle  \rangle_{\rm data}$ represents averages computed from the
data. The couplings inferred in this way can be thought of as some sort of
functional couplings between the neurons \cite{Shlens06,Yu08}.

The experimental studies by Schneidman {\em et al} and others
\cite{Shlens06,Tang08} showed that the pairwise Ising
model provided good approximations to the true distributions for
sets of up to $10$ neurons with small probability of spiking. These
experimental studies were followed by a theoretical analysis by Roudi {\em et al} \cite{Roudi09}
in which the authors studied the quality of pairwise models using a perturbative expansion
in $N\overline{\nu}\delta t$, where $N$ is the population
size, $\overline{\nu}$ is the mean population firing rate, and
$\delta t$ is the time bin chosen for binning the data. They showed that pairwise models always provide a
good statistical description of spikes as long as $N$ is small compared to 
$N_c=({\overline{\nu}\delta t})^{-1}$. That is, a pairwise model is
almost guaranteed to give a good result for any $N$
if the average firing rate is low enough and/or the time bins are small enough
such that $N\ll N_c$. This may be taken as a good news, but there is a  flip side to it:
finding that the Ising distribution is a good model for $N\ll N_c$ does
not tell us whether it is going to be a good model in the large $N$ limit.

In the first part of this paper, we address the first question
mentioned above, i.e., finding the parameters of the Ising model
given the measured means and correlations: the ``inverse Ising
problem.'' Here, we study various fast approximations for extracting
the couplings of the Ising model and compare their results with the
commonly used, usually slow, but potentially exact, Boltzmann
learning algorithm \cite{Ackley85}. We show that the couplings found using the
inversion of the TAP equations \cite{Kappen98,Tanaka98}, and the
approximation recently proposed by Sessak and Monasson
\cite{Sessak09} do a very good job in approximating true couplings.
Furthermore, the inversion of the TAP equations leads to
overestimating the couplings, while the Sessak and Monasson
approximation leads to underestimating them, and we note that simply
averaging them gives an even better result. We also study the
dependence of the inferred couplings on the size of the subset of
neurons for which the couplings are extracted. We find that the mean
and standard deviation of the couplings exhibit size dependences
compatible with those predicted for an infinite-range spin glass model (the SK model \cite{Sherrington75}) when it is
in its normal phase. We
also show that these dependences are mainly caused by scaling of the
individual couplings rather than restructuring the couplings.

We performed all this analysis for two sets of data. For one set,
which we describe as ``tonic firing'', the neurons in the network
fired at constant rates.  For the other set, which we call
``stimulus-driven'', the external input to the network was varied
temporally, evoking a modulation of the firing of all the neurons in
the network and, thus, additional ``stimulus-induced'' correlations.
Our findings are nearly same for the two data sets, so in most
of the following we show results only for the tonic case. The only
systematic difference between the results in the two cases is a
small increase in the mean of the inferred couplings in the
stimulus-driven case, as described in section \ref{tonicvsstim}.

All studies to date on the quality of the pairwise models have been
carried out for sets of neurons smaller than $N_c$ by factors of
$2-3$ (in the regime where a good pairwise fit is trivial
\cite{Roudi09}).  In section \ref{true vs Ising} we test the quality
of pairwise models for set sizes above $N_c$, using our data. We
find, as predicted in \cite{Roudi09}, that the fit quality
deteriorates as $N$ increases and this continues to be the case even for
$N > N_c$.

\section{Simulation Data} \label{sim_data}

We obtained our data from a simulated model cortical network of 1000
spiking neurons, 80\% of them excitatory and 20\% inhibitory,
operating in a high-conductance state \cite{Destexhe03} of
balanced excitation and inhibition \cite{Van98}.
There is a general consensus that cortical networks operate in such
a state.  The connectivity in the network was random, with a 10\%
probability of connection between any two neurons. The model is
fairly realistic: its neurons have Hodgkin-Huxley spike generating
dynamics, and its synapses are modeled as conducting ion channels
which are opened for short times by presynaptic spikes. Its membrane
potential and firing statistics are in good agreement with {\em in
vivo} measurements on local cortical networks. The details of the simulations
are described in Appendix \ref{sims_detail}.

Spike trains generated from the simulated network were divided into
bins of length $10$ ms. To each bin we then assigned a vector of
spin variables $\bS=(s_1,\dots,s_N)$ in which $s_i=-1$ if neuron $i$
did not emit a spike in that bin, and $s_i=1$ otherwise
\cite{Schneidman05,Shlens06,Tang08}. We then compute the mean
magnetization and pairwise correlations of these spin variables and
use them to fit the Ising model that generates the same mean and
pairwise correlations. In the analysis reported here, we only
studied excitatory cells with mean magnetization larger than $-0.98$
(i.e., firing rates greater than 1 Hz). We did this for two reasons.
First, the estimation of the means and, importantly, correlations
for cells with very small firing probabilities is very inaccurate.
Second, fitting these small numbers using essentially any method is
also inaccurate.

\section{Approximate and Exact Solutions to the Inverse Ising Problem}\label{approxandexact}

The simplest method for finding the fields and couplings such that
Eqs.~(\ref{means-corrs-p-data}) are satisfied is Boltzmann learning
\cite{Ackley85}. This is an iterative algorithm in which, at each
step, the fields and the couplings are adjusted as follows: starting
from some initial guess for the parameters, one computes the means
and the pair correlations under the Ising distribution using the
current values of the parameters. One then makes changes $\delta
h_i$ and $\delta J_{ij}$ in the parameters according to
\begin{subequations}
\begin{eqnarray}
&&\delta h_{i}=\eta \left\{\langle s_i \rangle_{\rm data}-
\langle s_i \rangle_{\rm Ising} \right\}\\
&&\delta J_{ij}=\eta \left\{\langle s_i s_j \rangle_{\rm data}-
\langle s_i s_j \rangle_{\rm Ising} \right\}.
\end{eqnarray}
\end{subequations}
One then recomputes the model means and correlations using the new
parameters, makes new parameter changes, and so forth until the
model statistics agree with the measured ones within the desired
accuracy.

The averages over the Ising distribution can be done either by exact
summation for small $N$, or by Monte Carlo sampling. In principle,
the Boltzmann learning is exact in the sense that it is guaranteed
to converge to the the correct fields and couplings after
sufficiently many minimization steps, and for sufficiently many
Monte Carlo steps per minimization step. Although Boltzman learning
is in principle exact, it is usually a slow algorithm. This is
particularly true for large $N$, for which one needs to run very
long Monte Carlo sampling steps per minimization step.  It is
therefore useful to have easily-computed approximate solutions,
either to use directly or as initial conditions for Boltzmann
learning. In this section, we compare the couplings obtained from
four fast approximation methods with those obtained from Boltzmann
learning.  The four approximations are (a) naive mean-field theory,
(b) an independent-pair approximation, (c) a combination of (a) and
(b) introduced recently by Sessak and Monasson (SM) \cite{Sessak09},
and (d) inversion of the Thouless-Anderson-Palmer (TAP) equations
\cite{Kappen98,Tanaka98} of spin-glass theory.  These four
approximations are described below.

\subsection{Naive Mean-Field Theory}

A naive mean-field theory (nMF) estimate can be derived simply by
differentiating the mean field equations for the magnetizations with
respect to the fields and using the fluctuation-response
relationship. Using the notation $m_i\equiv \langle s_i \rangle$ and
$C_{ij}\equiv \langle s_i s_j\rangle-m_im_j$, this yields
\begin{eqnarray}
C_{ij}&=&\frac{\partial m_i}{\partial h_j}=
\frac{\partial}{\partial h_j} \tanh(h_i+\sum_k J_{ik} m_k)\nonumber\\
&=&(1-m_i^2)\left[\delta_{ij}+\sum_{k} J_{ik} C_{kj}\right].
\label{MF}
\end{eqnarray}
Equivalently, one can write
\begin{equation}
{\sf J}^{\rm nMF}= {\sf P}^{-1}-{\sf C}^{-1} \label{J-MF}
\end{equation}
where $P_{ij}=(1-m_i^2)\delta_{ij}$.

\subsection{Independent-Pair Approximation}

One simple approximation is obtained by treating every pair of
neurons as if they were independent of the the rest of the system.
Consider two spins, $i$ and $j$, and let us denote the field on spin
$i$ ($j$) in the absence of the bond $J_{ij}$ between them by
$h^{c}_{i}$ ($h^{c}_{j}$). We can then write the probability
distribution over the states of this two-spin system as
\begin{equation}
Z_{ij}p_{s_i,s_j}=e^{h^c_is_i+h^{c}_js_j+J_{ij}s_is_j},
\label{Bethe-probs}
\end{equation}
where $Z_{ij}$ is the partition function of this two-spin system. In
writing the above equation, we assume that the state of spin $i$
will not have any effect on $h^{c}_j$ and vice versa. (A sufficient
condition for this to hold is that the system is on a Bethe
lattice.)

Eqs.~\ref{Bethe-probs} can be solved for $J_{ij}$:
\begin{equation}
J^{\rm Pair}_{ij}=\frac{1}{4}\log\left(\frac{p_{++}p_{--}}{p_{+-}p_{-+}}\right),
\label{hJ-p}
\end{equation}
where $p_{++}$ is the probability that both spins are up, $p_{+-}$ when the
first spin is up the second one is down etc. Expressing the probabilities in terms of the means and correlations.
one gets
\begin{eqnarray}
J^{\rm Pair}_{ij}=\frac{1}{4}\ln\left [\frac{(1+m_i+m_j+C^{*}_{ij})
(1-m_i-m_j+C^*_{ij})}{(1-m_i+m_j-C^*_{ij})(1+m_i-m_j-C^*_{ij})}\right], \nonumber\\
\label{pairs-appx}
\end{eqnarray}
where $C^*_{ij}=C_{ij}+m_im_j$.

In the low rate limit, $m_i\rightarrow -1$ and $m_j\rightarrow -1$,
the above expression simplifies to
\begin{equation}
J^{\rm LR}_{ij}=\frac{1}{4}\ln\left[
1+\frac{C_{ij}}{(1+m_i)(1+m_i)}\right], \label{lowrate-approx}
\end{equation}
which is identical to the result derived by perturbative expansion 
in $N\overline{\nu} \delta t$ in \cite{Roudi09} which we simply call
the low rate expansion. The fact that this low rate expansion
gives identical results  to the limiting case of independent pair approximation is
expected since for sufficiently low rates, the contribution of feedback loops to the local field on each
site can be neglected. Consequently, one can ignore the contribution
from other spins to the correlation function of $i$ and $j$.

\subsection{Sessak-Monasson Approximation}

Recently Sessak and Monasson \cite{Sessak09} derived an expression
relating the couplings to the means and correlations using a
perturbative expansion in the correlations. This was done by
extending the approach proposed by Georges and Yedidia
\cite{Georges91}: instead of performing one Legendre transform to
fix the magnetizations, as in \cite{Georges91}, they performed two
Legendre transforms of the free energy, one to fix the magnetization
and the other one to fix the correlations. They then expanded the
result in a high-temperature Plefka series \cite{Plefka82}. The
authors noticed that some of the terms in the expansion can be
summed up, yielding a closed form approximation for the couplings
that takes the form
\begin{equation}
J^{\rm SM}_{ij}=J^{\rm loop}_{ij}+J^{\rm Pair}_{ij}-\frac{C_{ij}}{(1-m^2_i)(1-m^2_j)-(C_{ij})^2},
\end{equation}
where $J^{\rm Pair}$ is given by Eq.~\ref{pairs-appx} and
\begin{equation}
J^{\rm loop}_{ij}=(L_iL_j)^{-1/2}\left[{\sf M}({\sf I}+{\sf
M})^{-1}\right]_{ij}
\end{equation}
with $L_{i}=1-m_i^2$, $M_{ij}=C_{ij}(L_{i}L_j)^{-1/2}$ and
$M_{ii}=0$. This expression for $J^{\rm loop}_{ij}$ can easily be
shown to be equivalent to the naive mean-field solution
Eq.~(\ref{J-MF}).

\subsection{Inversion of TAP Equations}

The TAP equations are mean-field equations that relate the local
magnetizations, $m_i$, to the external fields and the couplings
\begin{equation}
\tanh^{-1}m_i=h_i+\sum_{j} J_{ij}m_j-m_i\sum_j J_{ij}^2(1-m_j^2).
\label{TAP-eq}
\end{equation}
The right-hand side is the total internal field acting on spin $i$,
including the Onsager reaction field in the last term.
Differentiating Eq.~(\ref{TAP-eq}) with respect to $m_j$ and using
the fluctuation-response relation, one obtains
\begin{equation}
(C^{-1})_{ij}=\frac{\partial h_i}{\partial m_j} = -J^{\rm
TAP}_{ij}-2(J^{\rm TAP}_{ij})^2 m_i m_j, \hspace{0.5cm} (i\neq j)
\label{FD-eq}
\end{equation}

Given the means and correlations, we can solve Eq.~(\ref{FD-eq}) to
find the couplings and use the results in Eq.\ (\ref{TAP-eq}) to find
the external fields. This is the simplest version of the scheme
introduced by Kappen and Rodriguez \cite{Kappen98} and Tanaka
\cite{Tanaka98}. The TAP equations are exact in the limit of  ``infinite-range interactions" where 
the $J_{ij}$Õs have means and variances that scale like $1/N$. 
For arbitrary couplings, the TAP equations constitute the first two
terms in the Plefka series \cite{Plefka82} which is a small-coupling
(high-temperature) expansion. In principle, one can include
higher-order terms in the Plefka expansion instead of Eq.\
(\ref{TAP-eq}), compute its derivative to find the susceptibility and
use the fluctuation-response relation to relate it to the connected
correlations functions. Here we stop at the level of TAP equations.

\begin{figure}[]
\includegraphics[height=4.5 cm, width=8.5 cm]{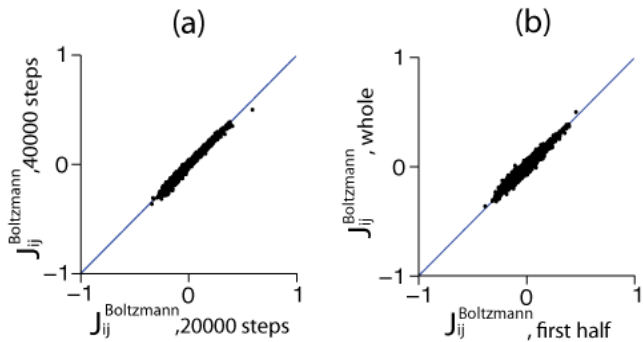}
\caption{Checking the reliability of the Boltzmann results. (a) The
couplings of a set of $N=200$ found using Boltzmann learning with
$20000$ learning steps versus those found after $40000$ learning
steps. (b) couplings learned from half of the data ($200000$ time
bins), versus those learned from all the data ($400000$ time bins).}
\label{HALFHB}
\end{figure}

\begin{figure}[]
\includegraphics[height=14 cm, width=8.5 cm]{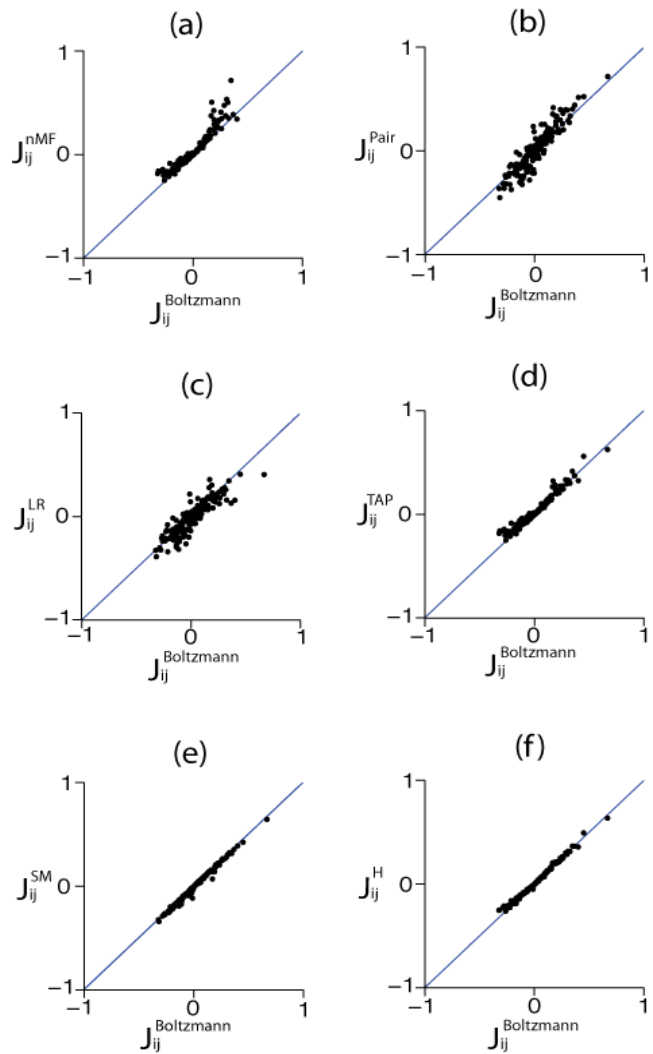}
\caption{Scatter plots comparing the solutions found from different
approximations with the Boltzman learning results for $N=20$. (a)
Naive mean-field approximation (b) Independent pair approximation, (c) low rate
limit, (d) TAP, (e) SM, and (f) a hybrid approximation obtained by
averaging TAP and SM.} \label{scatter20}
\end{figure}

\begin{figure}[]
\includegraphics[height=14 cm, width=8.5 cm]{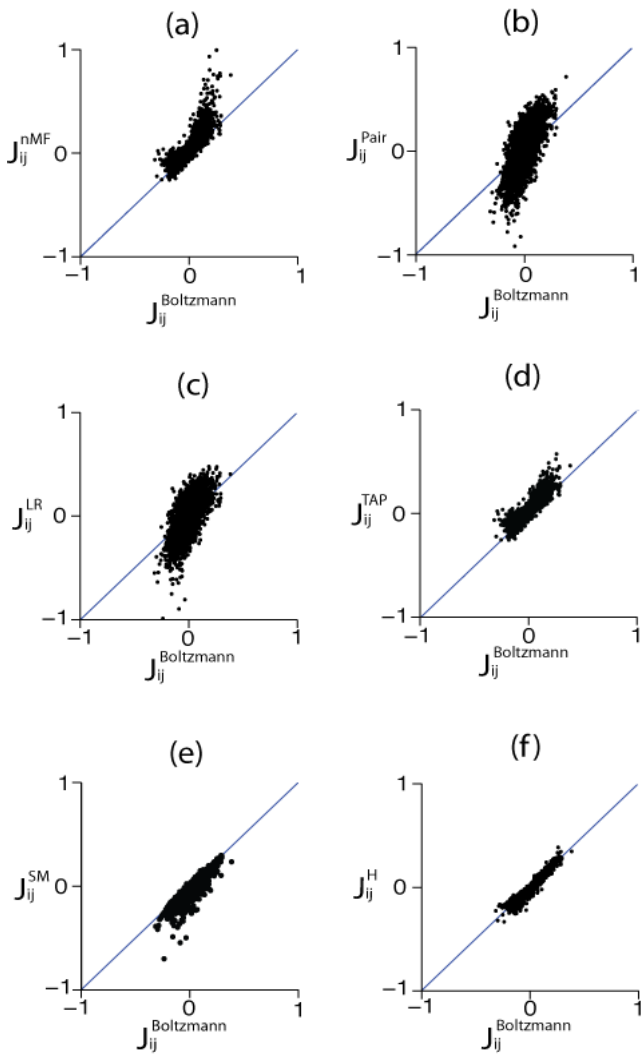}
\caption{The same as Fig. \ref{scatter20}, but for $N=200$ neurons.}
\label{scatter200}
\end{figure}

\subsection{Comparison between Boltzmann learning and the approximate solutions}

We considered $4000$ sec worth of data from our simulated network
(about $1$ hr, which is of the order of a stable retina recording
session), binned the data into 10-ms bins, and computed the means
and equal-time pairwise correlations of the spin representation of
the data (see sec. \ref{sim_data}). We first inferred the couplings
of the Ising model using $40000$ steps of Boltzmann learning with a
learning rate of $\eta=0.1$.  At each step, the model means and
correlations were computed on the basis of $30000$ Monte Carlo
sampling steps. We then compared these results with those obtained
from the approximation schemes listed in the preceding subsection. In this
comparison, we are assuming that the couplings inferred using
Boltzmann learning are the correct ones and judging the approximate
methods according to how well they agree with them. Before going on
with the comparison results, we take a moment to justify this claim.

The results of the Boltzmann learning may not be correct for two
reasons. First, the correlations passed to the Boltzmann algorithm
are, in general, not the true correlations, since they are computed
from finite data. Second, Boltzmann learning converges to the true
results only in the limit of infinitely many learning steps, and
there is always a chance that one has not run it long enough. To see
how much error such effects led to in our results, we conducted two
tests. First, we divided our spike trains of $4000$ sec into two
halves and computed two sets of correlations and means: one from the
first half of the data and other other from the whole set (this
latter set is what we use in the subsequent analysis). We then
scatter-plotted the $J_{ij}$'s inferred from the first half of the
data versus those computed from the full $4000$ seconds of
simulation.  We also plotted the results found after $20000$
learning steps against those found from $400000$ steps (using
correlations computed from all our data). The results are plotted in
Fig.\ \ref{HALFHB}. This figure shows that within the scale of the
errors of the various approximations (see Fig. \ref{scatter200}),
the Boltzmann results can be considered to be stable and accurate.

We now move on to the comparison of the couplings found from the
approximate solutions with those found from Boltzmann learning. The
results for a set of $N=20$ neurons is shown in Fig.\
\ref{scatter20}. This figure shows that for this subset size, all
approximations do well, although the Sessak-Monasson
approximation outperforms the others by a small margin, followed
closely by the TAP-inversion solution. For larger sets, the
difference between the quality of different approximations becomes
more clear. Fig.\ \ref{scatter200} shows the results for $N=200$.
Here the Sessak-Monasson approximation and TAP inversion outperform
the rest by a significant amount. Figs.\ \ref{scatter20} and
\ref{scatter200} also show that the SM and the TAP-inversion
approximations differ in the way their errors: SM tends systematically
to underestimate the couplings, while TAP inversion overestimates
them. This suggests that naively averaging the two, i.e., summing
them and dividing by two, should do a better job. The results of
such a hybrid approximation shown in Figs.\ \ref{scatter20}f and
\ref{scatter200}f confirm this expectation.

\begin{figure}[h]
\includegraphics[height=11 cm, width=8.5 cm]{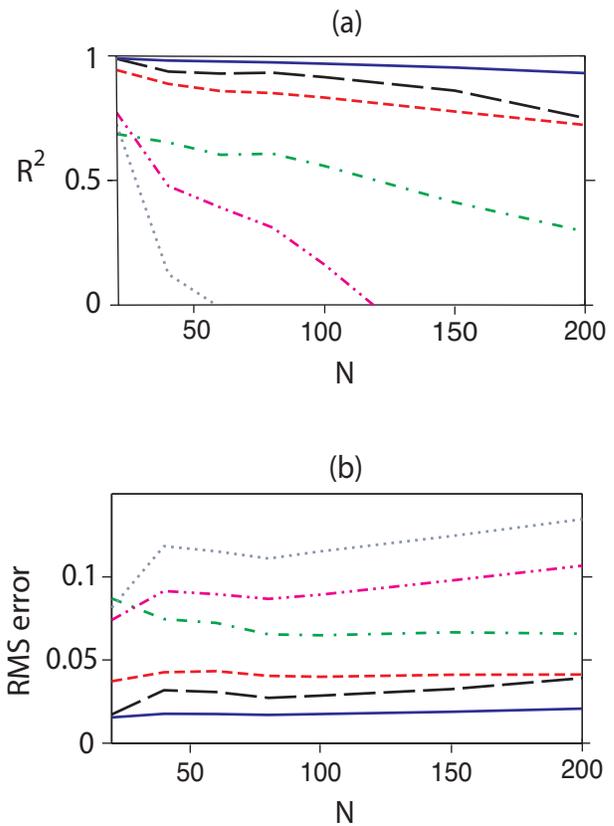}
\caption{Quantifying the performance of different approximation by
computing $R^2$ (defined in Eq.\ \ref{Rsq}) and the RMS error
(defined in Eq.\ \ref{RMSe}) between the approximate solutions and
the result of Boltzmann learning, as a function of $N$.  Black (long
dashed line), SM; Red (short dashed line), TAP; Blue (full curve),
hybrid SM-TAP; Green (dashed dotted line), nMF; Magenta (dashed
double dotted), low-rate approximation; Gray (small dotted),
independent-pair approximation.} \label{summary-perf}
\end{figure}

\begin{figure}[h]
\includegraphics[height=14 cm, width=8.5 cm]{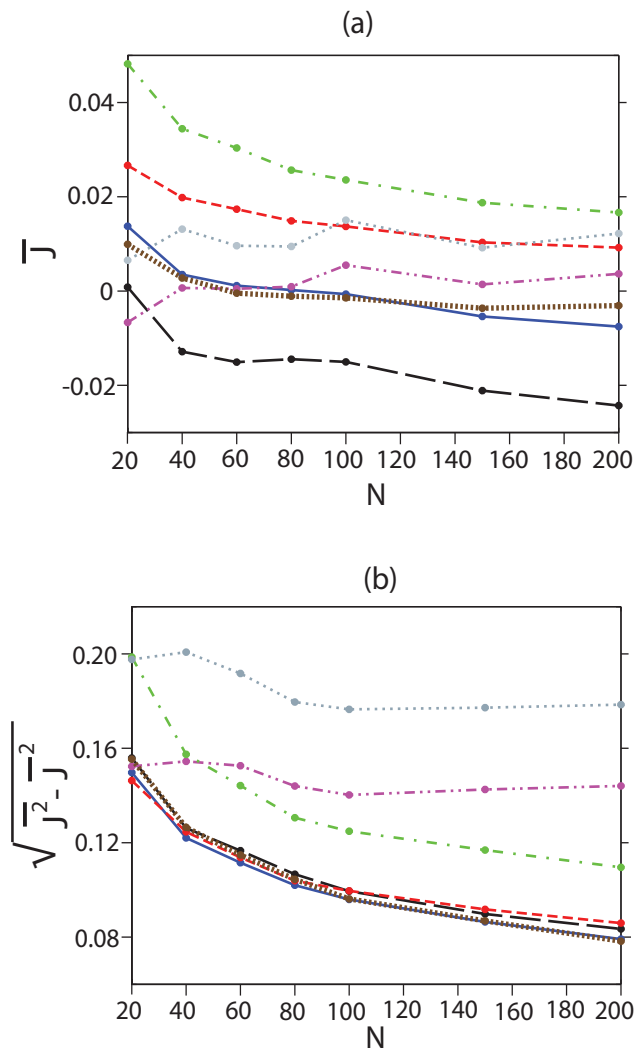}
\caption{The $N$-dependence of the mean and standard deviation of
the solutions found from different approximations and Boltzmann
learning.  Black (long dashed line), SM; Red (short dashed line),
TAP; Green (dashed dotted line), nMF; Blue (full curve), hybrid SM-TAP; 
Magenta (dashed double dotted), low-rate approximation; Gray
(small dotted), independent-pair approximation; Brown (large dots),
Boltzmann.} \label{mean-std-N}
\end{figure}

The performance of different approximations as a function of set
size is shown in a more systematic way in Fig.\ \ref{summary-perf}.
In this figure, we computed the similarity between the Boltzmann
solution and the approximate ones for different sizes, using two
quantities as measures of similarity. The first was the coefficient
of determination, $R^2$, defined as
\begin{equation}
R^2\equiv 1-\frac{\sum_{ij} (J^{\rm approx}_{ij}-J^{\rm
Boltzmann}_{ij})^2}{\sum_{ij} (J^{\rm
Boltzmann}_{ij}-\overline{J^{\rm Boltzmann}_{ij}})^2}, \label{Rsq}
\end{equation}
where $J^{\rm approx}_{ij}$ refers to the couplings inferred using
one of the aforementioned approximations, and $\overline{J^{\rm
Boltzmann}_{ij}}=\sum_{i\neq j} J^{\rm Boltzmann}_{ij}/(N(N-1)$.
Values of $R^2$ close to one indicate good approximations. We also
considered the RMS error defined as
\begin{equation}
\sqrt{\frac{1}{N(N-1)}\sum_{i\neq j} (J^{\rm approx}_{ij}-J^{\rm
Boltzmann}_{ij})^2}. \label{RMSe}
\end{equation}
As shown in Fig.\ \ref{summary-perf}, TAP inversion, SM and the hybrid
TAP-SM outperform the other approximations for all $N$ and according to both measures.

We also studied the relationship between the $N$-dependence of the
couplings inferred using Boltzmann learning with those found from
the approximate solutions. The results are shown in Fig.\
\ref{mean-std-N}.  The standard deviation of the $J_{ij}$'s is
well-approximated by both TAP inversion and the Sessak-Monasson
formula (and therefore also by their average).  Naive mean-field theory captures
the decrease with $N$ qualitatively correctly but gives an estimate
which is systematically too large. The independent-pair
approximation fails (as does its low-rate limit,
Eq.~(\ref{lowrate-approx})). We will study the $N$-dependence of the couplings  
more carefully in the next section.

The means of the $J_{ij}$'s are smaller than their standard
deviations by an order of magnitude or more.  The Boltzmann value is
indistinguishable from zero for $N > 50$.  Of all the
approximations, only the SM-TAP average seems to approximate it
well, although SM, TAP and naive MF estimates all show a decrease
with $N$.

If averaged over many samples, the independent-pair approximations
will not exhibit any $N$-dependence in either means and standard
deviations of the couplings.  Thus, they can not capture the
observed systematic decreases of these statistics with $N$.

\begin{figure}[h]
\includegraphics[height=15 cm, width=8.5 cm]{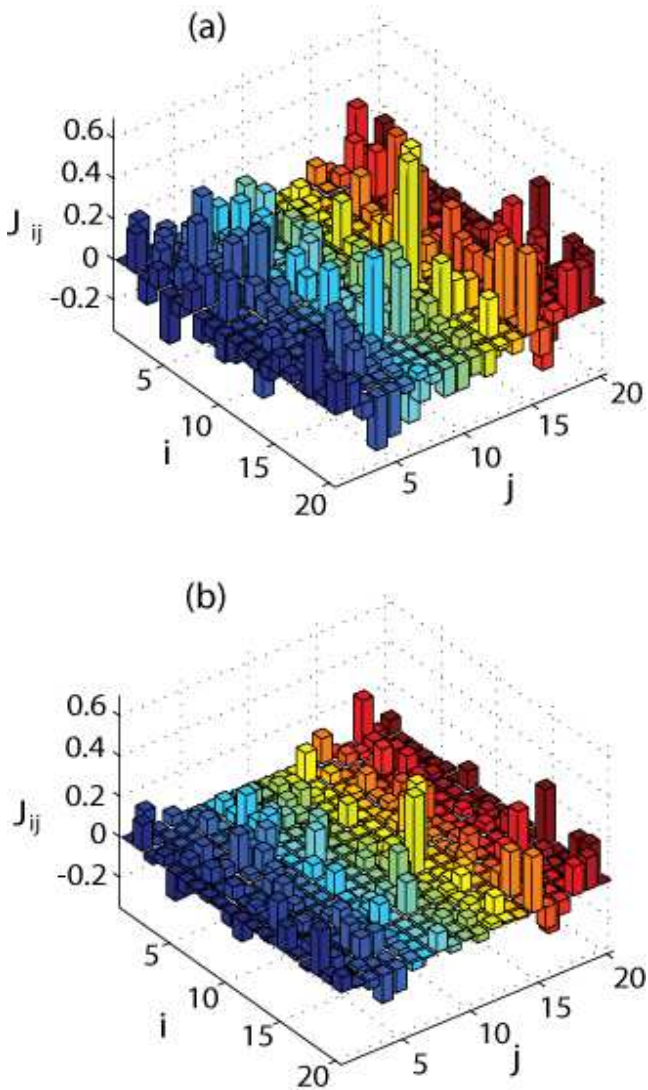}
\caption{The couplings among $20$ neurons, inferred when no
additional neurons are considered (a), and when inferred as a part
of a network of size $200$ (b). This figure shows that the
$N$-dependence of the mean and standard deviation of the couplings
in Fig.\ \ref{mean-std-N} arises from the scaling of individual
couplings.} \label{scaling20200}
\end{figure}

\begin{figure}[h]
\includegraphics[height=6 cm, width=8.5 cm]{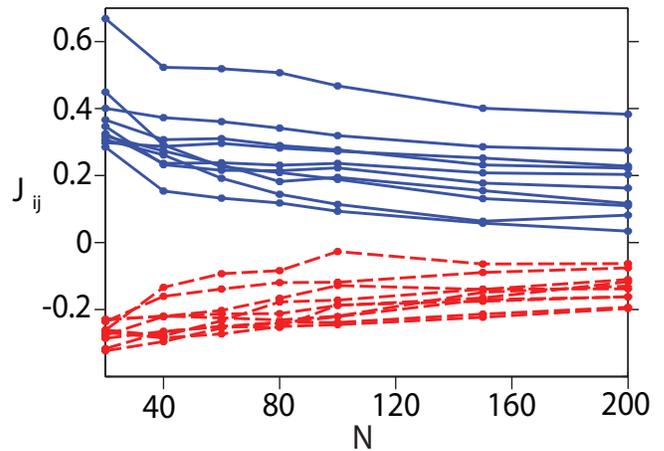}
\caption{The behaviour of the 10 largest and 10 smallest couplings
in the set of 20 neurons, when more and more neurons are added. The
blue (full) curves show how the 10 largest weights in the population
of 20 neurons change their values as we ass more and more cells, and
the red (dashed) curves show the same thing for the 10 smallest
couplings. The weights are inferred using Boltzmann learning.}
\label{10_small_10_big}
\end{figure}

\begin{figure}[h]
\includegraphics[height=13.5 cm, width=8.3 cm]{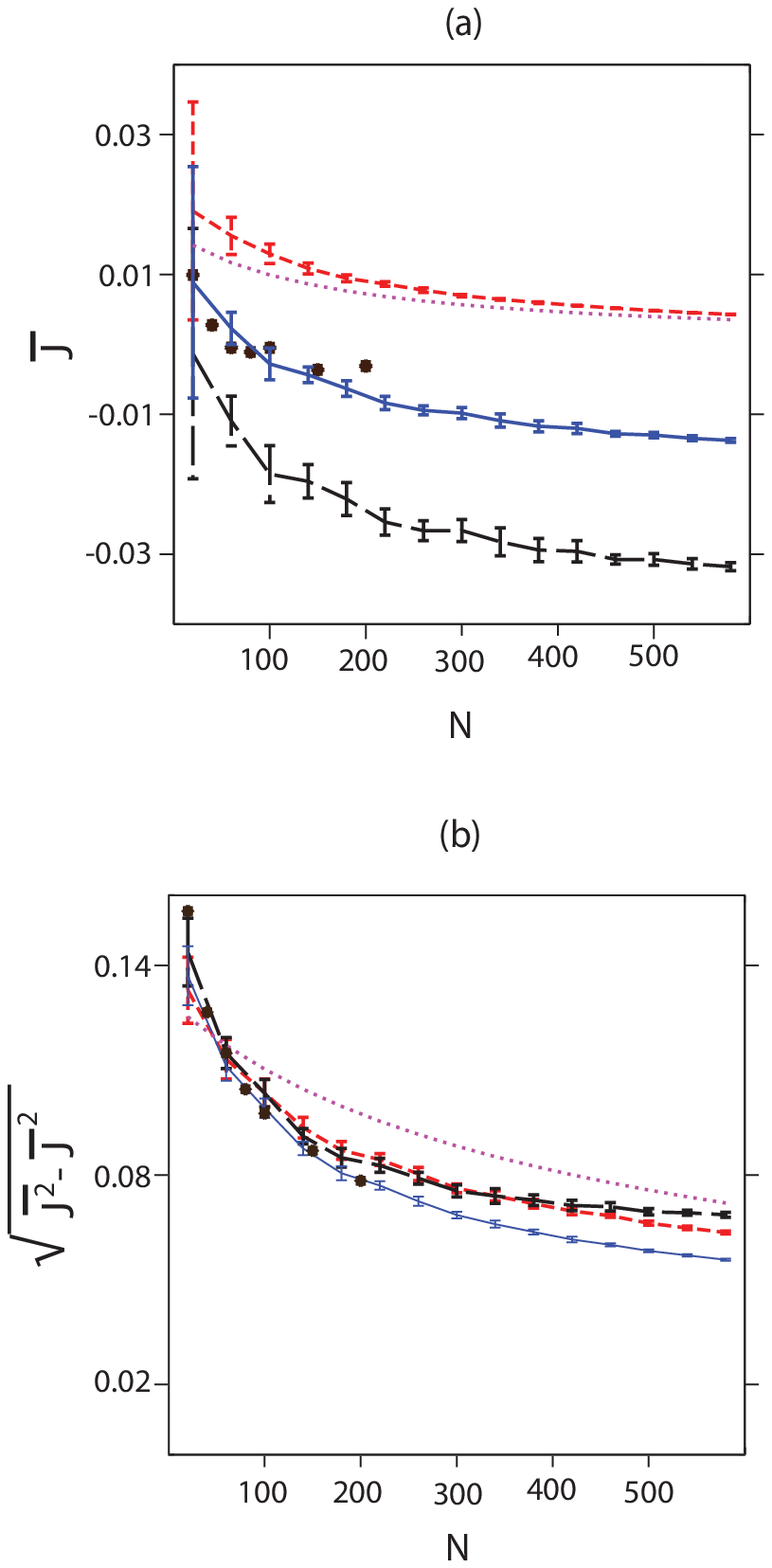}
\caption{Comparing the $N$-dependence of the mean and standard
deviation of the solutions found from the first three good
approximations  (SM (black, long dashed), TAP (red, short dashed),
SM-TAP hybrid (solid blue)), with the SK  prediction
(Magenta, dotted). The Boltzmann results for individual samples up
to $N=200$ are replotted from Fig.\ (\ref{mean-std-N}) for
comparison (brown stars).} \label{N_dep}
\end{figure}

\section{Scaling of inferred couplings and the comparison
with mean-field spin glass behavior} \label{scaling}

Fig.\ \ref{mean-std-N} shows that the mean and standard deviation of
the inferred couplings have some sort of scaling behaviour with
population size and that this behaviour is well preserved in the
TAP, SM and hybrid TAP-SM approximations. This $N$-dependence could be either
due to partial or complete restructuring of the couplings when more
and more neurons are added, or simply due to the scaling of
individual couplings. Fig.\ \ref{scaling20200} and Fig.\
\ref{10_small_10_big} show that what is actually happening is the
latter. Fig.\ \ref{scaling20200}a shows the inferred couplings
between $20$ neurons when no additional neurons are considered,
while Fig.\ \ref{scaling20200}b shows the couplings between these
neurons when they are extracted from the data of a set of $200$
neurons. As one can see in this figure, the main structure of the
couplings of the subset is retained in the larger set, but they are
scaled down. In Fig.\ \ref{10_small_10_big}, we trace how the
largest ten (full blue lines), and the smallest (i.e., most
negative) ten (red dashed lines) of a set of $20$ neurons change as
we add more and more neurons to the pool. As can be seen, the small
weights increase their values, and the large weights decrease their
values with $N$, but only the weights that are very close to each
other cross.  This observation suggests that the structure of the
weights is preserved as $N$ is increase, i.e. there is an
approximate scaling behaviour for individual weights.

How can we explain this scaling of the weights? Because there is no
spatial structure in the connectivity of our original simulated
model, we do not expect to find any such structure in the functional
connections $J_{ij}$ that we obtain. Thus, it seems possible that
our inferred models will be  ``infinite-ranged.''
If so, we may describe the statistics of the couplings by the infinite-range
Sherrington-Kirkpatrick (SK) model of a spin glass
\cite{Sherrington75}, in which one assumes independently distributed
$J_{ij}$ with a mean $J_0/N$ and a variance $J^2/N$.
This model both has a normal and a spin glass phase, The normal phase may be characterized
completely in terms of two order parameters, the mean magnetization
$M = N^{-1}\sum_i m_i$ and the mean square magnetization $q =
N^{-1}\sum_i m_i^2$.  In the spin glass phase, a much more complex
description is required. Fortunately, as shown below, we will only need to
consider the normal phase, where one can derive relatively simple
relations that express the mean and variance (across all pairs of
spins) of the correlations in terms of the mean and variance of the
couplings and the order parameters of the model. These relations can
be inverted to give the mean and variance of the $J$s in terms of
the mean and variance of the $C$'s. The size $N$ of the system
enters in these relations, so they make predictions about the
$N$-dependence of the $J$s that we can compare with the results of
the inference algorithms for different sizes of sets of neurons.

Consider first the average (over off diagonal elements) of the
correlation matrix, $\overline{C} = [N(N-1)]^{-1}\sum_{i \neq j}
C_{ij}$.  Standard mean-field arguments (e.g., averaging
Eq.~(\ref{MF})) lead to
\begin{equation}
\overline{C} = \frac{J_0(1-q)^2/N}{1 - J_0(1-q)}. \label{Cbarresult}
\end{equation}
The mean square of the correlation matrix elements can also be shown
to be \cite{deAlmeida78} (see also \cite{Fischer93} sec. 3.2)
\begin{equation}
\overline{C^2} = \frac{J^2 S^2 /N}{1 - J^2S},  \label{varCresult}
\end{equation}
where
\begin{equation}
S = \frac{1}{N}\sum_i (1-m_i^2)^2 = 1 - 2q + N^{-1}\sum_i m_i^4.
\label{Sdef}
\end{equation}

Inverting Eqs.~(\ref{Cbarresult}) and (\ref{varCresult}) to obtain
the statistics of the $J$s in terms of those of the $C$s, we find
\begin{eqnarray}
\overline{J} = \frac{J_0}{N} = \frac{\overline{C}}{(1-q)(1-q + N
\overline{C})} \label{meanJresult} \\
{\rm var}(J) = \frac{J^2}{N} = \frac{\overline{C^2}}{S(S + N
\overline{C^2})}  \label{varJresult}
\end{eqnarray}

These results hold for $N$ equal to the full system size.  However,
in the inference process described here, one works with data from a
smaller number of neurons and tries to model them by a network of
that smaller size.  Thus, the inferred $J$s will have (larger) means
and variances than their true ones because the $N$ in the
denominators of Eqs.~(\ref{meanJresult}) and (\ref{varJresult})
will be smaller than the true size.  (Note that the statistics of
the measured $C$s do not, on average, depend on the size of the set
of neurons.)

The criterion for the stability of the normal phase \cite{deAlmeida78} is $J^2S > 1$.  Using our 
extracted values of $J$ and calculating $S$ from Eq.~\ref{Sdef}, we find
that $J^2S$ grows with $N$ but never exceeds $0.65$.  Therefore the assumption that the
system is in the normal phase is self-consistent.  

To the extent that our inferred network is like an SK model,
Eqs.~(\ref{meanJresult}) and (\ref{varJresult}) should describe the
way the mean and variance of the $J$s very with $N$.  This is easy
to test because the statistics of the $C$'s and the moments of the
$m_i$ that occur in $q$ and $S$ are readily calculated from the
spike data. Fig.\ \ref{N_dep} shows how well the results of the
inference algorithms conform to this simple behavior. In this
figure, the means computed from the SM, TAP, and hybrid SM-TAP
approximations, averaged over $20$ random samples of the excitatory
neurons for $N<300$ and $10$ samples for $N>300$, are shown,
together with the results of Boltzmann learning on individual
samples and the mean-field spin glass predictions. The quantities
$S, q, \overline{C}$ and $\overline{C^2}$ that appear in Eqs.\
(\ref{varJresult}) and (\ref{meanJresult}) are computed from the
whole population of excitatory neurons.

The agreement of Eqs.~(\ref{meanJresult}) and (\ref{varJresult})
with the results of our parameter extraction methods is not perfect,
but the magnitude of the standard deviation of the $J_{ij}$'s and
its falloff with $N$ are captured reasonably well.  The mean agrees
well with the TAP result, but of course the TAP $J_{ij}$'s are
systematically higher than the true (Boltzmann) ones, as described
above.

\section{Tonic versus Stimulus-driven Firing States}
\label{tonicvsstim}

\begin{figure}[h]
\includegraphics[height=15 cm, width=8.5 cm]{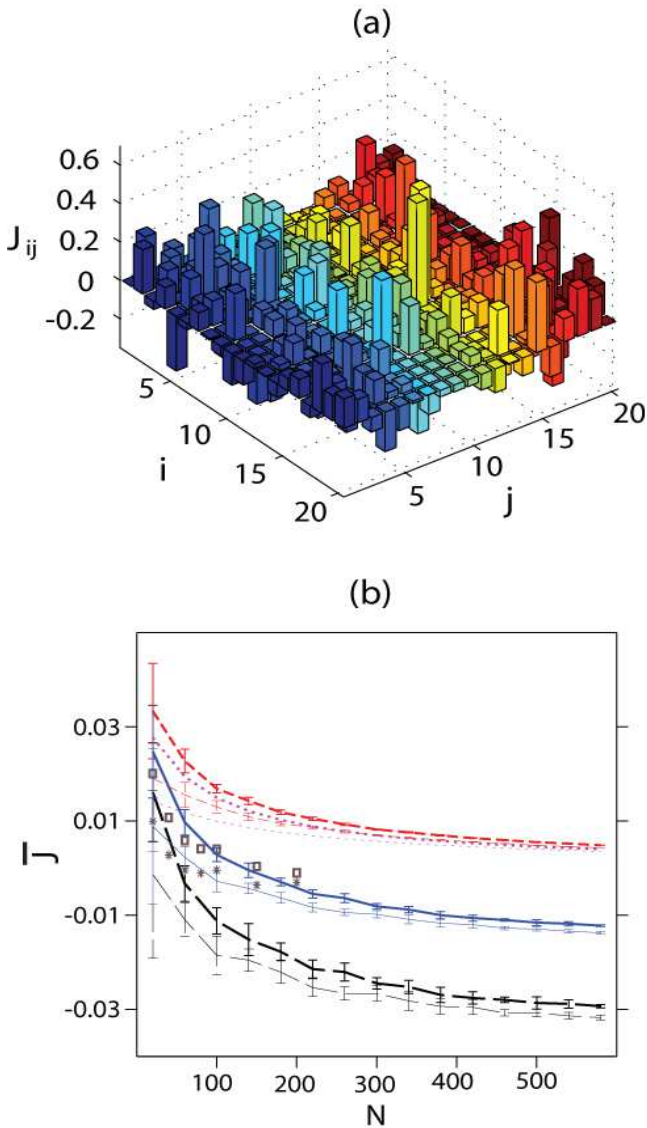}
\caption{The $N$-dependence of the mean couplings in the
stimulus-driven case as found from Boltzmann learning on single
samples (brown squares), SM (black long dashed lines), TAP (red
short dashed lines), SM-TAP hybrid (blue full curve),  and the
prediction of Eq.\ (\ref{meanJresult}) (magenta dotted line). The
thick curves are the results for the stimulus-driven case, and the
results of the tonic case are plotted using thin curves.}
\label{Mean_N_600_Tonic_Stim}
\end{figure}

In the previous sections, we studied tonic-firing data only.  When
we conducted the same analyses for the stimulus-driven data, all the
same conclusions about the quality of different approximations and
their size dependencies were drawn. The only difference that we
observed is  slightly higher mean coupling found in the
stimulus-driven case. The similarity between the weights can be seen
by comparing Fig.\ \ref{Mean_N_600_Tonic_Stim} and Fig.\
\ref{scaling20200}a.  The increase in the couplings of the weights
is too small to see here.  It is also too small to show up in a
scatter plot.

The shifts in the mean coupling computed by Boltzmann learning and
the good approximation methods (SM, TAP and their average) are shown
in Fig.\ \ref{Mean_N_600_Tonic_Stim}b. While only the SM-TAP average
gets the mean coupling right, all three methods capture the
stimulus-induced shift.

The higher mean couplings can be understood qualitatively using
Eq.~\ref{meanJresult}.  In the stimulus driven case, the mean
correlation is slightly higher than the tonic case (the difference
is what is generally called ``stimulus-induced correlations''). For
the data used for Fig.\ \ref{Mean_N_600_Tonic_Stim}b, $\overline{C}_{\rm
stim}=0.0052$ versus $\overline{C}_{\rm tonic}=0.0023$, leading in
Eq.~(\ref{meanJresult}) to a larger $\overline{J}$.

\section{The True Distribution Versus the Model
distribution}\label{true vs Ising}

In this section, we study the second question about pairwise models raised in the 
introduction, i.e. the quality of the model
in approximating the true distribution of spikes.
To compare the fitted Ising distribution, $p_{\rm Ising}$, with the
true distribution, $p_{\rm true}$, we considered the
Kullback-Leibler (KL) divergence \cite{Kullback51} between the two,
defined as
\begin{eqnarray}
D_{\rm KL}(p_{\rm true}||p_{\rm Ising})&=&\sum_{\bS} p_{\rm true}(\bS) \ln\left(\frac{p_{\rm true}(\bS)}{p_{\rm Ising}(\bS)}\right).\nonumber\\
&\equiv& d_{\rm Ising}
\label{KL-ising}
\end{eqnarray}
In addition, we considered the KL divergence between the true
distribution and an independent-neuron distribution
\begin{equation}
p_{\rm ind}\propto \exp\left\{\sum_{i}h^{\rm ind}_i s_i\right\},
\label{p-ind}
\end{equation}
in which the external fields $h^{\rm ind}_i$ are chosen so that Eqs.\
\ref{means-p-data} are satisfied. We also define $d_{\rm ind}\equiv
D_{\rm KL}(p_{\rm true}||p_{\rm ind})$. Denoting the averages of $d_{\rm Ising}$ and
$d_{\rm ind}$ over many samples of a given size $N$ by  $\overline{d}_{\rm Ising}$ and $\overline{d}_{\rm ind}$, we can define
\begin{equation}
G\equiv 1-\frac{\overline{d}_{\rm Ising}}{\overline{d}_{\rm ind}}
\label{G-def}
\end{equation}
as a measure of the goodness of pairwise models
\cite{Schneidman05,Shlens06,Tang08, Roudi09}. (Other studies
\cite{Tang08,Roudi09} have used the measure $\Delta = \overline{d}_{\rm
Ising}/\overline{d}_{\rm ind} = 1-G$.) It is easy to show that
\begin{subequations}
\begin{align}
&&d_{\rm ind}=S_{\rm ind}-S_{\rm true}\\
&&d_{\rm Ising}=S_{\rm Ising}-S_{\rm true},
\end{align}
\end{subequations}
and consequently
\begin{equation}
G= \frac{\overline{S_{\rm ind}-S_{\rm Ising}}}{\overline{S_{\rm ind}-S_{\rm
true}}},
\end{equation}
where $S_{\rm Ising}, S_{\rm ind}$ and $S_{\rm true}$ are the
entropies of $p_{\rm Ising}, p_{\rm ind}$ and $p_{\rm true}$, and the overline
indicated averaging over many samples of the same size. The quantity $G$
is the fraction of the entropy difference between the independent
model and the data that is explained by the pairwise model. When $G$ is near one, the pairwise models is very
good (compared to the independent model) in terms of the amount of
the true entropy that it explains. When $G=0$, the pairwise model is
just as bad as the independent-neuron model.

\begin{figure}
\includegraphics[height=18 cm, width=8.5 cm]{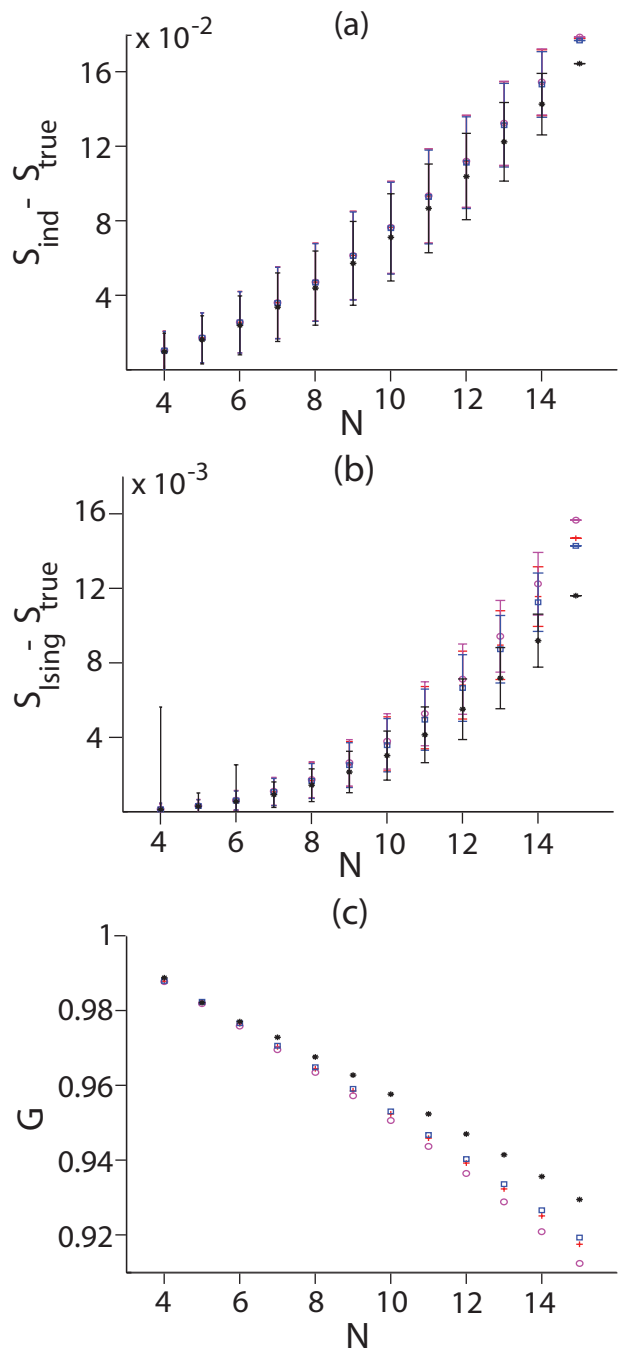}
\caption{The behavior of the KL distances and $G$ versus the
set size, $N$. (a) $d_{\rm ind}$, the KL distance between the
independent and the true distributions versus $N$, (b) $d_{\rm
Ising}$, the KL distance between the Ising and the true distribution
versus $N$, (c) the goodness of fit,
$G$, versus $N$. Magenta circles, red crosses and blue
squares represent estimations of these quantities from $T=10^6$,
$T=1.5\times 10^6$ and $T=1.8\times 10^6$ samples respectively,
while black stars show the bias-corrected ($T\rightarrow \infty$)
estimates.} \label{Ents}
\end{figure}

In Fig. \ref{Ents}, we show the behavior of $d_{\rm ind}$, $d_{\rm
Ising}$ and $G$ versus $N$. To produce this figure, we first chose a
population of $15$ neurons. For each $N$, we then chose ${15 \choose
N}$ or $2500$ random populations of $N$ neurons from the original
$15$ cells, whichever was larger. For each of these populations, we
computed the entropy using $T$ time bins from the simulation by
simply counting the number of occurrences of each pattern. We also
computed the means and correlations from these $T$ time bins and fit
an independent-neuron and an Ising model to the data. Fitting of the
parameters of the Ising model was done by numerically exact
minimization of the error function $\sum_{{\bS}} p_{\rm true}(\bS)
\log[p_{\rm true}(\bS)/p_{\rm Ising}(\bS)]$, using conjugate
gradient descent. We then computed the entropies of the independent
and Ising models using brute force summation over all the states. We
did this procedure using $T=10^6$, $T=1.5\times 10^6$ and
$T=1.8\times 10^6$. The resulting values of $d_{\rm ind}, d_{\rm
Ising}$ and $G$ for each of these values of $T$ is shown in Fig.\
\ref{Ents}. As expected, finite $T$ leads to an underestimation of
$S_{\rm true}$ and thus overestimation of $d_{\rm ind}$ and $d_{\rm
Ising}$. To correct for finite sampling bias, the resulting values
of $d_{\rm ind}$ and $d_{\rm Ising}$ for each $T$ were fit by a
second order polynomial in $1/T$, and the limit $T\rightarrow
\infty$ was taken \cite{Strong98}. The unbiased estimates are shown
in black in Fig.\ \ref{Ents}. Both $d_{\rm ind}$ and $d_{\rm Ising}$
increase with $N$, while $G$ decreases. For the population shown
here we had $N^{-1}\sum_i \langle s_i \rangle_{\rm data}\approx
-0.8$, indicating that $N_c \approx 10$. This figure shows that for
small populations $G$ is close to $1$, but it decreases linearly
even for values of $N$ above $N_c$.

\section{Discussion}

Even with their shortcomings, models of the type we have studied
here provide a potentially attractive framework for analyzing
multi-neuron spike data.  We imagine that experimentalists would
want a quick and easy way to find out what $J_{ij}$Õs characterize
the spike data they have measured.  In previous work, the extraction
of the $J_{ij}$Õs was done by brute force, using Boltzmann learning.
This is in principle exact but very slow for large $N$.   The fast,
approximate parameter-extraction methods described here offer a way
to make Ising pair models a practical data-analysis tool.

We calibrated these fast methods by comparing their results with
that of Boltzmann learning for sets of neurons up to a size $N =
200$, which took several days for a single run.   We were able in
this way to evaluate and compare several methods: (1) the
independent-pair approximation, which is known \cite{Roudi09} to be
correct in the low-firing rate, small-$N$ limit, (2) inversion of
the mean-field equations for the correlations, (3, SM) a combination
of (1) and (2) proposed by Sessak and Monasson \cite{Sessak09}, and
(4, TAP) inversion of the Thouless-Anderson-Palmer equations.

Of these approximations, the third and fourth turned out to work
very well, with SM slightly better that TAP. SM has a slight
tendency to underestimate the $J_{ij}$Õs and TAP has a slight
tendency to overestimate them.  We found that an ad hoc averaging of
the SM and TAP  $J_{ij}$'s agreed even better with the Boltzmann
learning results, with an rms error of about half that achieved by
SM or TAP. Of course, this result has to be taken as just a bit of
good luck for the particular network used to generate these data;
there is no reason to expect it to hold generally.

We could then proceed to apply the good fast approximation schemes
for $N > 200$ and to identify the important generic features of the
extracted couplings.  We found that for larger $N$ the $J_{ij}$'s
had a nearly zero mean value and that their absolute values appeared
to shrink systematically as $N$ increased. Furthermore, the
$J_{ij}$'s found to be strongest (i.e., to have the largest absolute
values) for one set of neurons were also generally found to be the
strongest within that set when the extraction was done for a larger
set of neurons. Thus, the strong $J_{ij}$'s appeared to be quite
robust statistics.

A measure of the typical magnitudes of the $J_{ij}$'s is provided by
their standard deviation.  Although the fit is not perfect, the
decrease in the standard deviation with $N$ is captured crudely by a
simple theoretical picture in which one assumes that the $J_{ij}$'s
are chosen randomly and independently. In other words, as far as its
pair correlations are concerned, the network behaves approximately
like a Sherrington-Kirkpatrick (infinite-range) spin glass. This finding
is not too surprising, in view of the fact that the network used to
generate the data had completely random connectivity.   Perhaps it
is more surprising that the data deviate systematically from the
spin-glass prediction.  We do not have any explanation of these
deviations.

We analyzed spike data for both tonic-firing and stimulus-driven
conditions.  In the latter, the input to the simulated cortical
network was varied temporally.  The fact that all neurons received
input with the same temporal modulation might be expected to
generate extra correlations, which would be reflected in increased
$J_{ij}$'s. In fact, this did happen, but the effect was very small.
The values of the $J_{ij}$'s found in the two conditions were nearly
the same.  Thus, the couplings obtained (in particular, the ways
they vary from one pair or neurons to another) are intrinsic
properties of the system.  The only systematic effect of the
stimulus was a small increase in the average $J_{ij}$.   This
weakness of this effect is perhaps to be expected, because the
temporal modulation employed was rather slow (a time constant of 100
ms) compared to response times in the network (10 ms or less).   It
would of course be of interest to study the effect of varying the
time constant of the input rate fluctuations.

A natural question to ask is whether the $J_{ij}$'s one finds have
any relation to the synaptic connections in the network that
generated the data from which they are extracted. In our case, we
know that synaptic connectivity, so we can answer this question.
Somewhat disappointingly, however, we have found no significant
relation between the $J_{ij}$'s and the synapses.  We believe this
is because we are trying to force a description with symmetric
couplings ($J_{ij} = J_{ji}$) onto a network where this symmetry is
absent or nearly so.  We think another approach would be required to
uncover the underlying synaptic connectivity, one based on
correlation between spikes by different neurons in different time
bins, rather than, as here, coincidences in the same time bin.  Then
one might be in a position to identify which neurons' spikes tend to
cause spikes in which other neurons, which is more closely related
to synaptic connectivity.

Of course, even within the present paradigm, pair models are not
guaranteed to be exact.  For our data and for small subsets of the
neurons ($N \le 15$), we were able to quantify the degree of
mismatch in terms of the Kullback-Leibler distance between the true
distribution and the Ising model Gibbs distribution. In agreement
with earlier results \cite{Schneidman05,Shlens06,Tang08,Roudi09}, we found
that pairwise Ising models perfectly model the true distribution in the limit of small $N$.
We also found that the quality-of-model measure, $G$ (Eq.~\ref{G-def}), decreases
linearly with $N$ for the range that we tested. For $N\ll N_c $, this decrease can be
understood using the expansion in $N\overline{\nu}\delta t$ of Roudi {\em et al}
\cite{Roudi09}. To lowest order one has $d_{\rm ind}\propto (N\overline{\nu}
\delta t)^2$ and $d_{\rm Ising}\propto (N\overline{\nu} \delta t)^3$; consequently
$G\propto 1-N\overline{\nu} \delta t$. However, this expansion is bound to break down,
as it will eventually predict a true entropy that decreases with $N$. This can be seen by
noting that $S_{\rm ind}\propto N$ and therefore, $S_{\rm true}=S_{\rm ind}-d_{\rm ind}=
c_1 N- c_2 N^2$ will be a decreasing function of $N$ for $N>c_1/(2c_2)$.  Nevertheless,
$G$ can still be a decreasing function of $N$ even when the expansion breaks down, and
indeed we see in Fig.\ \ref{Ents} that this is the case for our data. The decrease in our data
is of the order of $5\%$ for $N = 10$, suggesting that one should be cautious in applying
pair models for $N$ bigger than about $50$ or so (where the pair model only explains about
75\% of the entropy difference between the independent-neuron model and the data). If we
naively extrapolate the linear dependence of $G$ on $N$, we find $G \approx 0$ for
$N \approx 200$, indicating that at this size a pairwise model would be no improvement on an
independent-neuron model.

Nevertheless, even when they are not good models (in the sense that
the Kullback-Leibler distance between them and the true distribution
is not small), pairwise models offer a conceptually simple and
useful framework for characterizing measured multi-neuron spike
statistics.  One the one hand, by construction, they describe the
first- and second-order statistics correctly, and on the other hand
they are the only models for which it is practically feasible to
carry out the fit using data sets of realistic size.  When used with
caution, they can provide robust and reliable information about the
correlation structure in the data, and the fast approximations we
have described here should be useful in applying them in practical
data analysis.

\appendix
\section{The Simulated Model Cortical Network}
\label{sims_detail}
The means and correlation functions we use in this paper are
obtained from simulating a network consisting of $1000$
Hodgkin-Huxley-like model neurons with conductance-based synapses,
$800$ excitatory and $200$ inhibitory. The network was driven by an
external population of $800$ excitatory neurons. The connectivity
was random, with the probability of a synapse between any two
neurons equal to $10 \%$. There was no synaptic randomness beyond
that implied by the random connectivity: When a synapse was present,
it had a maximum conductance which depended only on which
populations the pre- and postsynaptic neurons belonged to. The
dynamics of the membrane potential, $V_{ia}$, of a neuron $i$ in
population $a= E$ (excitatory) or $I$ (inhibitory) is given by
\begin{eqnarray}
\frac{dV_{ia}}{dt}= && -\sum_{\sigma} G^{\sigma,{\rm
intr}}_{ia}(t)(V_{ia}-V^{\rm rev}_{\sigma})\nonumber
\\&-&\sum_{bj} G^{\rm syn}_{ia,bj}(t) (V_{ia}-V^{\rm rev}_b),
\end{eqnarray}
where $V^{\rm rev}_{E}=0$ and $V^{\rm rev}_{I}=-80$ mV,
$G^{\sigma,{\rm intr}}_{ia}$ is the intrinsic conductance of type
$\sigma =$ Leak, Na, or K of neuron $i$ in population $a$, and
$G^{\rm syn}_{ia,bj}(t)$ is the conductance associated to the
synapse from neuron $j$ in population $b$ to neuron $i$ in
population $a$. The intrinsic conductances are of the standard
Hodgkin-Huxley form $G^{\sigma,{\rm intr}}=g^{0}_{\sigma}
m_{\sigma}^{p_\sigma} h_{\sigma}^{q_\sigma}$, with $p_{\rm Na}=3$,
$q_{\rm Na}=1$, $p_{\rm K}=4$, $q_{\rm K}=0$, $p_{\rm Leak}=q_{\rm
Leak}=0$. The gating variables $m_{\sigma}$ and $h_{\sigma}$ obey
standard kinetics with voltage-dependent opening and closing rates.
The forms of these rates, as well as the values of the
$g^0_{\sigma}$, were taken from  and Par{\'e}
\cite{Destexhe99}.

The synaptic conductances $G^{\rm syn}_{ia,bj}(t)$ were obtained by
filtering the presynaptic spike trains through a sequence of three
exponential filters.  The time constants of these filters,
representing the synaptic delay, the rise time and the fall time of
the synaptic conductance after a presynaptic spike, were chosen
randomly from uniform distributions of means 1, 3, and 5 ms, and
half-widths equal to 90\% of the means, respectively.

In the tonic state the firing rate of the external population was
constant, leading to constant firing rates for the neurons in the
network. Because of the randomness in the network structure, these
rates varied from neuron to neuron. The maximum synaptic
conductances were chosen so that the total average synaptic
conductance was in the range found by Destexhe {\em et al}
\cite{Destexhe03} and so that the inhibitory neurons fired on
average at about twice the average rate of the excitatory ones. In
the stimulus-driven state, the rate of the external population was
modulated randomly in time, in order to mimic qualitatively the
experiments of Schneidman {\em et al} \cite{Schneidman05}, where
movies of dynamic natural scenes were projected onto salamander
retinas. Specifically, we took the external rate to be a constant
plus exponentially filtered white noise, with a time constant of
$100$ ms. As a result, the firing rates of the neurons in the
network also varied in time. The noise parameters were chosen so
that the averages of the firing rates over intervals much longer
than the time constant were approximately the same as those in the
tonic state.

\bibliography{mybibliography}
\end{document}